\documentclass[11pt]{JHEP3}
\usepackage{palatino}
\usepackage{amsmath}\usepackage{amssymb}

\author{Joakim Munkhammar\\ Studentstaden 23:230, 752 33,
Uppsala, Sweden\\
E-Mail: \email{joakim.munkhammar@gmail.com}}

\title{Is Holographic Entropy and Gravity the result of Quantum
Mechanics?}


\abstract{In this paper we suggest a connection between quantum
mechanics and Verlinde's recently proposed entropic force theory
for the laws of Newton. We propose an entropy based on the quantum
mechanical probability density distribution. With the assumption
that the holographic principle holds we propose that our suggested
quantum entropy generalizes the Bekenstein entropy used by
Verlinde in his approach. Based on this assumption we suggest that
Verlinde's entropic theory of gravity has a quantum mechanical
origin. We establish a reformulation of the Newtonian potential
for gravity based on this quantum mechanical entropy. We also
discuss the notion of observation and the correspondence to
classical physics. Finally we give a discussion, a number of open
problems and some concluding remarks.}

\begin{document}


\section{Introduction}

%
%
%
%

In a remarkable paper Verlinde recently proposed a framework for
gravity as an entropic force \cite{Verlinde}. This theory, while
related to Jacobsson's approach \cite{Jacobsson} and subsequent
work by Padmanabhan \cite{Padmanabhan,Padmanabhan2,Padmanabhan3},
showed that Newtonian gravity easily could be obtained by using
entropic and holographic arguments. The assumption was that space
is emergent and that the holographic principle holds
\cite{Verlinde}. Bekenstein entropy was also a key component in
his approach. He thus reversed the line of research, assuming that
the holographic principle was underlying Newtonian physics
\cite{Smolin}. The change of entropy was linked to the change of
the Newtonian potential, this led to the conclusion that inertia
might be equivalent to the lack of entropy gradients
\cite{Verlinde}. As Verlinde states, the holographic principle has
not been easy to extract from the laws of Newton and Einstein
because it is deeply hidden within them \cite{Verlinde}. His paper
attracted quite some attention and several papers from various
fields of theoretical physics, including Loop Quantum Gravity and
quantum mechanics, have been published relating to its topic
\cite{Ghosh,Smolin,Vancea,Zhao}. A shortcoming of the theory was
the unknown origin of the coupling constant $\hbar$
\cite{Bekenstein,Verlinde}. This coupling constant was added by
Bekenstein \cite{Bekenstein} in the 1970s mainly for dimensional
reasons and has since remained a mystery. We will suggest an
origin of this constant in this paper.



\section{Entropy and the holographic principle}
In Verlinde's view space is mainly a storage place of information,
which is associated with positions, movements and mass of matter
\cite{Bousso,Padmanabhan,Padmanabhan2,Susskind,tHooft,Verlinde}.
This information is displayed to us on a surface, a holographic
screen. The information is stored in discrete bits on the screen
and since the number of bits is limited we get holographic
effects. This means that if there is more information on the
inside than the amount of information accessible on the screen
then information will be hidden from us as we observe the
dynamics. This is the holographic principle. Thus the dynamics on
the screen is governed by some unknown rules which then only can
utilize the information on the screen. Since information is stored
on a screen this means that space is emergent in the normal
direction of the screen. The microstates may be thought of having
all sorts of physical attributes such as energy, temperature etc.
This is then related, via entropy, to the information associated
with the system \cite{Verlinde}.

\subsection{Entropy as a force}
Bekenstein related the area of a black hole to the entropy of it
by assuming that all information lost down a black hole must still
be conserved and therefore contained in some measure
\cite{Bekenstein,Hawking}. This laid the foundation for an
emergent holographic view of physics \cite{Verlinde}. The
connection between entropy and information is that the change of
information $I$ is the negative change of entropy $S$:
\begin{equation}\label{Information}
\Delta I = - \Delta S.
\end{equation}
Considering a small piece of a holographic screen and a particle
with mass $m$ approaching it from the side at which time has
already emerged, then Verlinde concluded (unitilzing Bekensteins
arguments) that the entropy associated with this process should be
Bekenstein entropy with an extra factor of $2\pi$ \cite{Verlinde}:
\begin{equation}\label{EntropicForce}
\Delta S = 2 \pi k_B \frac{mc}{\hbar} \Delta x.
\end{equation}
Here $k_B$ is Boltzmanns constant and the factor $2\pi$ was added
for reasons to be clear in connection with the gravitational
force. Furthermore, an entropic force is a macroscopic force that
originates in a system with many degrees of freedom by the
universe's statistical tendency to maximize its entropy. An
entropic force $F$ is defined as \cite{Verlinde}:
\begin{equation}\label{Force}
F \Delta x = T \Delta S
\end{equation}
where $T$ is temperature. In order to relate the entropy to the
screen the maximum number of bits $N$ that can be associated with
a screen is then assumed to be:
\begin{equation}\label{NumberofBits}
N = \frac{A c^3}{G \hbar} = \frac{4 \pi R^2 c^3}{G \hbar},
\end{equation}
where $A = 4\pi R^2$ is the area of the screen. The temperature
can be determined from the equipartition rule
\eqref{EntropicForce} \cite{Padmanabhan,Padmanabhan2}:
\begin{equation}\label{BitEnergy}
E = \frac12 N k_B T,
\end{equation}
which is the the average energy per bit. We shall also assume the
mass-energy relation:
\begin{equation}\label{EMC}
E = Mc^2.
\end{equation}
In a straight forward way these equations yields the gravitational
force:
\begin{equation}
F = G \frac{Mm}{R^2}.
\end{equation}
This is a surprising result considering it practically comes from
first principles. In addition to this Verlinde also discusses the
nature of inertia and via the equipartition rule for single a
particle he gets:
\begin{equation}
mc^2 = \frac12 n k_B T.
\end{equation}
Here $n$ is the number of bits associated with a particle. He
associated this with the Unruh effect:
\begin{equation}\label{Unruh}
k_B T = \frac{1}{2 \pi} \frac{\hbar a}{c}
\end{equation}
where $a$ is the acceleration, which can be set equal to the
gradient of the Newtonian potential:
\begin{equation}
a \equiv \nabla \Phi.
\end{equation}
From this Verlinde derives the following relation \cite{Verlinde}:
\begin{equation}
\frac{\Delta S}{n} = - k_B \frac{a \Delta x}{2c^2} = - k_B
\frac{\Delta \phi}{2 c^2}.
\end{equation}
A general conclusion here is that the change of the Newtonian
potential $\phi$ is related to the change of entropy $S$. Verlinde
continued and generalized these concepts to a relativistic version
of this entropic gravitational theory which for strong fields
turns out to be equivalent to Einstein field equations, see
\cite{Jacobsson,Verlinde} for more information. His general
conclusion is that inertia is due to the lack of entropy
gradients, and conversely that gravity is due to the presence of
them \cite{Verlinde}.



\section{Quantum mechanics and entropy}
Quantum mechanics has historically a number of different but
equivalent approaches \cite{Lisi}. The perhaps most canonical is
the path-integral formulation made by Feynman \cite{Feynman,Lisi}.
A Quantum mechanical wave function $\psi$ is linked to the
classical action $A$ via the relation:
\begin{equation}
\psi = R e^{i\frac{A}{\hbar}},
\end{equation}
where $R^2 = \psi \psi^\dagger = |\psi|^2$ is the probability
density distribution ($\psi^\dagger$ is the complex conjugate of
$\psi$) \cite{Bransden}. Feynman's insight was that any quantum
mechanical system is the sum of all complex amplitudes relating to
a particles paths from one point to another \cite{Feynman}. This
meant in practice:
\begin{equation}\label{QMWave}
\psi = R e^{i\frac{A}{\hbar}} = \sum_n e^{i\frac{A_n}{\hbar}},
\end{equation}
where the sum goes over all actions of the possible paths of a
particle from one point to another. It has many similarities with
the partition function of statistical mechanics \cite{Lisi}. We
have the definition of the action as the time-integral over the
lagrangian:
\begin{equation}
A_n = \int \mathcal{L}_n \,dt. 
\end{equation}
The probability density for a particle is defined as:
\begin{equation}\label{Rho}
\rho = \psi \psi^\dagger = |\psi|^2,
\end{equation}
where $\psi^\dagger$ is the complex conjugate of $\psi$. When
integrating \eqref{Rho} one gets the probability of a state in a
particular domain of space. We shall here assume that the
integration be over the volume $V_\mathcal{S}$ inside a given
holographic screen gives unity:
\begin{equation}\label{Normalization}
\int_{V_\mathcal{S}} |\psi|^2 dV = 1.
\end{equation}
This excludes the possibility of the particle being outside the
screen. The application of this assumption will be apparent when
we discuss the nature of observation and correspondence to
classical physics in section \ref{Observation}. Via the Feynman
approach \eqref{QMWave} to quantum mechanics we can conclude that
$|\psi|^2$ is related to a sum of states of a quantum system. In
the single particle situation it contains information regarding
the probability of position of the particle. In light of this we
suggest that the probability density $|\psi|$ is in fact related
to a partition function $Z$ for different possible states:
\begin{equation}\label{Partition}
Z \equiv |\psi|^{-2}.
\end{equation}
Although the particular sum of the partition function is not known
it is no stretch of imagination to assume that at least one exists
for every physical system, especially considering the structure of
the Feynman approach to quantum mechanics \eqref{QMWave}. If we
assume that the partition function \eqref{Partition} holds we can
construct an entropy:
\begin{equation}\label{GeneralEntropy}
S = k_B \frac{\partial (T ln(Z))}{\partial T} = k_B \frac{\partial
(T ln(|\psi|^{-2}))}{\partial T},
\end{equation}
which shall be referred to as the \textit{quantum entropy}
associated with a physical system. In those cases where $ln(Z)$
are independent of $T$ in some way we have a simplified entropy:
\begin{equation}\label{SpecialEntropy}
S = k_B ln(Z) = - 2 k_B ln|\psi|.
\end{equation}
This relation between $|\psi|$ and $S$ will be used as a
generalization of the Bekenstein entropy throughout the rest of
this paper. Note that the entropy here is space dependent.

\subsection{Single stationary particle entropy}
Lets take the particular case of a single particle at rest to see
how the entropy works. We have the Klein-Gordon equation from
relativistic quantum mechanics as follows \cite{Bransden}:
\begin{equation}\label{KleinGordon}
\Big(\nabla^2 -\frac{1}{c^2} \frac{\partial}{\partial t}\Big) \psi
= \frac{m^2 c^2}{\hbar^2} \psi,
\end{equation}
which for a stationary particle becomes:
\begin{equation}\label{KGStationary}
\nabla^2 \psi = \frac{m^2 c^2}{\hbar^2} \psi.
\end{equation}
This equation has the solution (that is square integrable):
\begin{equation}\label{WaveSingle}
\psi(x) = A e^{-\frac{mc}{\hbar}x},
\end{equation}
where $A$ is a normalization constant and we consider $x$ the
radius outwards from the classical position of the particle. Note
that in the case where $x$ is considered the radius a very small
potential $\hbar c/r$ will be apparent in both the Schr\"odinger
and Klein-Gordon equations, it is a potential which in most
situations can completely be omitted. If we insert
\eqref{WaveSingle} into \eqref{GeneralEntropy} we get:
\begin{equation}\label{AnotherEntropy}
S = k_B \frac{\partial T ln|\psi|^{-2}}{\partial T} = -2 k_B
\frac{\partial T (-\frac{mc}{\hbar}x + ln(A))}{\partial T},
\end{equation}
here we shall have to use the temperature from the equipartition
law (to be used and derived in section \ref{Newtonian}):
\begin{equation}\label{AnotherTemp}
T = \frac{\hbar G M}{2k_B c x^2}
\end{equation}
for an external mass $M$. If we evaluate \eqref{AnotherEntropy} we
get:
\begin{equation}\label{AlmostEntropy}
S = - 2 k_B \Big(-\frac{mc}{\hbar}x + ln(A)\Big).
\end{equation}
Note that when using the particular temperature
\eqref{AnotherTemp}, which is canonical here, and a $|\psi|$ that
is exponentially decaying, most of the time (like in this case)
the special entropy solution \eqref{SpecialEntropy} will work. We
shall therefore use the special entropy solution from here on. Now
if we return to \eqref{AlmostEntropy} and look at the difference
in entropy $\Delta S = S_1-S_2$ from the difference in $\Delta x =
x_1-x_2$ we get:
\begin{equation}\label{delta}
S_1 - S_2 = \Delta S = 2k_B \frac{mc}{\hbar} x_1 - ln(A) - (2k_B
\frac{mc}{\hbar} (x_2) - ln(A))
\end{equation}
The $\ln(A)$-terms disappear and the negative sign vanishes for
the right choice of direction on the difference $\Delta x$ making
\eqref{delta}:
\begin{equation}\label{EntropyAlmostVerlinde}
\Delta S = 2k_B \frac{mc}{\hbar}\Delta x
\end{equation}
which is equivalent to the entropy used in Verlinde's approach
\eqref{EntropicForce} up to a factor of $\pi$. Verlinde added the
$2\pi$-factor for the convenience of cancelling it in the pursuit
of the gravitational force \cite{Verlinde}. In order to remedy
this problem we redefine the proportionality constant between the
number of bits on the screen and the entropy by substituting
$\hbar \to \pi \hbar$:
\begin{equation}\label{NumBits2}
N \equiv \frac{A c^3}{\pi \hbar G}
\end{equation}
This move ought to be considered legal since the proportionality
\eqref{NumberofBits} was estimated \cite{Bekenstein,Verlinde}.
Thus we have concluded that in this framework we get the
expression for Bekenstein's entropy in the situation of a
stationary particle. In order to make sense of what entropy is in
this approach we shall apply it to Verlinde's approach to gravity
in section \ref{Newtonian}.




\section{Newtonian gravity from quantum mechanics via
entropy}\label{Newtonian} In light of his discovery that entropy
might be the source of gravity, Ted Jacobsson stated that a
quantization of general relativity is physically as absurd as the
quantization of for example the wave equation for sound in air
\cite{Jacobsson}. In a similar fashion we shall not quantize
gravity here, but rather construct gravity based on the quantum
mechanical entropy which is analogous to the case of the sound
waves in air where the underlying microstates are quantum
mechanical and the macroscopical wave is derived from them. Lets
take the equation for entropic force:
\begin{equation}\label{EntropicForce2}
F \Delta x = T \Delta S
\end{equation}
and for infinitesimal displacements we have the integral form:
\begin{equation}\label{EntropicForceDiff}
U = \int F dx = \int T dS.
\end{equation}
The potential energy is the result of the integral of the
temperature over entropy in the emergent direction (normal to the
screen). We assume that the particle subject to force creating the
potential energy $U$ has mass $m$ and that this potential energy
can be defined as the product of a potential $\phi$ and the mass
$m$:
\begin{equation}
U = m \phi
\end{equation}
The temperature $T$ on the screen in \eqref{EntropicForce2} is
defined via the equipartition rule of all energy inside a
holographic screen:
\begin{equation}
E = Mc^2 = \frac12 k_B n T,
\end{equation}
and if we utilize the area to entropy relation \eqref{NumBits2}
and rearrange for $T$ we get:
\begin{equation}
T = \frac{\hbar G M}{2k_B c r^2},
\end{equation}
where $r$ is the radius of the screen. This is the temperature of
the screen at radius $r$. With the aid of our entropy defined from
quantum mechanics \eqref{GeneralEntropy}:
\begin{equation}
S = -2k_B ln|\psi|,
\end{equation}
we may re-express the entropy expression
\eqref{EntropyAlmostVerlinde} on a differential formulation:
\begin{equation}\label{EntropyGradient}
d S = -2 k_B d (ln|\psi|) = -2 k_B |\psi|^{-1} d |\psi|.
\end{equation}
If we insert \eqref{EntropyGradient} in \eqref{EntropicForceDiff}
we arrive at a general potential energy $U$ for a particle:
\begin{equation}
U = m \phi = -\int \frac{\hbar GM}{c r^2} d (ln|\psi|) = -\int
\frac{\hbar GM}{cr^2 |\psi|}d |\psi| = \frac{\hbar}{c} \int
\frac{\nabla \phi_N}{|\psi|} d |\psi|,
\end{equation}
where $\phi_N$ is the Newtonian potential. A generalized potential
energy emerges as $U$. Note here that this for the single particle
can be regarded as a generalization of the potential:
\begin{equation}\label{NablaPhi}
\phi = \frac{\hbar}{mc} \int \frac{\nabla \phi_N}{|\psi|} d |\psi|
\end{equation}
If we take the special case of the single particle solution
\eqref{WaveSingle} and insert it into \eqref{NablaPhi} terms
cancel and we get:
\begin{equation}
m \phi = m \phi_N = G \frac{Mm}{r}
\end{equation}
which is the Newtonian potential energy, just as in Verlinde's
approach \cite{Verlinde}. Also, the gradient of the force is
naturally identified with the gradient of the Newtonian potential
in this special case:
\begin{equation}
\nabla \phi = \nabla \phi_N = -\frac{GM}{r^2}
\end{equation}
for which we have Poisson's equation:
\begin{equation}\label{Poisson}
\nabla^2 \phi = 4 \pi G \rho.
\end{equation}
Thus we have derived Poisson's equation for gravity via the
stationary solution of a single particle \eqref{WaveSingle}. The
expression \eqref{Poisson} can also be reformulated as Gauss's
law:
\begin{equation}
M = \frac{1}{4\pi G} \int_{\mathcal{S}} \nabla \phi \cdot dA.
\end{equation}
We have the general potential (which is a generalization of the
Newtonian potential) \eqref{NablaPhi}:
\begin{equation}
\phi = \frac{\hbar}{mc} \int \frac{\nabla \phi_N}{|\psi|} d
|\psi|.
\end{equation}
It should be stressed here that this potential is the potential
that the single particle with mass $m$ experiences, so in reality
the potential is directly coupled to the particular force of the
particle. This reformulation of the Newtonian potential could be
used in any quantum theory as an additional potential in order to
incorporate gravity up to some generalized Newtonian limit. The
Hamiltonian $\mathcal{H}$ should be supplied with the extra term
containing the potential in order to contain gravity:
\begin{equation}\label{Hamiltonian}
\mathcal{H} \to \mathcal{H} + \int T dS =  \mathcal{H} +
\frac{\hbar}{c} \int \frac{\nabla \phi_N}{|\psi|} d |\psi|.
\end{equation}
If we assume that we have a particle $m$ in a potential with a
massive mass $M$ as above in \eqref{Hamiltonian} then we can
insert it in for example the Schr\"odinger equation
\cite{Bransden} which gives:
\begin{equation}
i \hbar \frac{\partial}{\partial t} = -\frac{\hbar^2}{2m} \nabla^2
\psi + \int T dS = -\frac{\hbar^2}{2m} \nabla^2 \psi
+\frac{\hbar}{c} \Bigg( \int \frac{\nabla \phi_N}{|\psi|} d |\psi|
\Bigg) \psi.
\end{equation}
Although we used the relativistic approach with the Klein-Gordon
equation in order to construct the single stationary particle
solution this equation should give interesting results within some
reasonable limits. Note that the gravitational potential that
arises here is only due to the entropic force acting on a
particle, so gravity is not added, only the entropic force. We
shall leave the relativistic approach to this quantum gravity for
future research.

\section{Observation and the correspondence
principle}\label{Observation} Our quantum approach to entropy
suggests that information in a physical system is directly
associated with its quantum mechanical entropy defined by the
equality of the partition function with the inverse probability
density distribution:
\begin{equation}
Z \equiv |\psi|^{-2}.
\end{equation}
How does this relate to physical observation? If we condense the
notion of observation in physics we can conclude that:

\vspace{10pt}

\begin{center}
\textit{"Observing a physical system is obtaining information from
it"}
\end{center}

\vspace{10pt}

This asserts that an observer observes a system. In the
holographic scenario we might imagine the observer being on the
outside of a screen observing it. When observing a physical system
it means that the entropy of the system will change because
information is obtained on it. This information needs to originate
from inside the screen and travel via some mediating particle, a
photon for example, and transmit information to a detector outside
the screen. In practice this means that the screen for the
particle, when the particle is observed, has a very small radius.
By the normalization condition:
\begin{equation}
\int_{V_\mathcal{S}} |\psi|^2 dV = 1,
\end{equation}
where $V_{\mathcal{S}}$ is the volume of the screen, we can see
that as $V_{\mathcal{S}} \to 0$ we have that:
\begin{equation}\label{PsiDelta}
|\psi|^2(\vec{x}) \to \delta(\vec{x}-\vec{x}_0).
\end{equation}
Here $\delta(\vec{x}-\vec{x}_0)$ is the Dirac delta function where
$\vec{x}_0$ is the point where the particle is detected. One
should keep in mind that the entropy is related to the screen
should be seen as a difference:
\begin{equation}
\Delta S = S(r) - S(0).
\end{equation}
This means that entropy approaches zero for the particle as the
radius of the sphere goes to zero (when observation takes place):
\begin{equation}
V_\mathcal{S} \to 0 \Rightarrow |\psi| \to \delta \Rightarrow
\Delta S \to 0
\end{equation}
Note here that even if $\Delta S \to 0$ it does not mean that
$\partial_r S \to 0$ since $\partial_r S$ is constant in for
example the free single stationary solution \eqref{WaveSingle}.
This entire situation is the equivalent to a wave function
collapse.


\subsection{Correspondence principle}
The correspondence to classical physics is when all screens either
become transparent (displaying all information) or collapse down
to single points. This happens for example when $\hbar \to 0$:
\begin{equation}
\lim_{\hbar \to 0} N = \lim_{\hbar \to 0} \frac{A c^3}{\pi \hbar
G} = \infty.
\end{equation}
As all holographic effects vanish the probability becomes binary
for each event for every observer. This means that the wave
function $|\psi|$ approaches the delta function \eqref{PsiDelta}
and all entropy vanishes. Generally one may conclude that in the
classical scenario all entropy vanishes. A particular effect of
this is that if all entropy vanishes, so does all quantum- and
gravitational effects. That all quantum effects vanishes in the
classical situation is not surprising, but that all gravitational
effects vanish is quite remarkable.

%

\section{Discussion}
\subsection{Quantum mechanics, information and entropy}

\subsubsection{The validity of quantum entropy}
The interpretation of the quantum mechanical uncertainty as a form
of entropy is perhaps not that strange considering that entropy in
fact is practically equivalent to lack of information regarding an
object. A main uncertainty in the approach proposed in this paper
is the connection between the proposed quantum entropy arising in
quantum mechanics and the thermodynamical entropy. In defence of
this assumption, the entropy in quantum mechanics must in some
way, for sure, be accounted for in the complete
thermodynamical-entropy theory of physics. Especially if we
consider the close information-to-physics connection arising from
the holographic principle and black hole thermodynamics. Also, the
relation between entropy and information hints that entropy should
be additive in the final correct setup of physical entropy:

\begin{equation}
\Delta S = \sum_i \Delta S_i = - \sum_i \Delta I_i
\end{equation}

which coupled with the indestructibility of information suggests
that at least one term is quantum mechanical in origin, this
provides a non-zero lowest estimate on the entropy contribution
from quantum mechanics. In fact other types of contributions from
quantum mechanics have been proposed recently \cite{Lee}.

\subsubsection{The single particle solution}
Our particular single particle solution \eqref{QMWave} might be
considered to be a special case given that it derives primarily
from the Klein-Gordon equation, which does not hold for all types
of particles. As far as free particle solutions goes the most
reasonable probability density solutions that can be normalized
will be those of exponential monotonic decreasing form. Whether it
be linear or non-linear, the dynamics will be similar to the one
proposed here, at least quantitatively. However there might be
particular qualitative difference for certain situations, these
are situations which are reasonably typical quantum-dominated
situations such as paricles in a box etc.

\subsubsection{Prospects}
An interesting aspect is that the principle law of the universe to
minimize entropy gradients via the second law of thermodynamics
will surely be important for the continued development of the
relation between quantum mechanics and Verlinde's theory. The
junction of thermodynamics, quantum mechanics and relativity has
not yet been fully understood, but it appears to have promising
prospects for future research via Verlinde's reversal of physics
through the holographic foundation for Newtonian mechanics.


\subsection{Conclusions}
The great connections between matter and information made by
Bekenstein, Hawking and others in the 1970s has turned out to have
very interesting consequences. Verlinde's framework for the origin
of the laws of Newton including gravitation based on entropy is
perhaps one of the greatest consequences of this. In what way
space is emergent and how the holographic principle holds is
starting to fall in to place. In this paper we have proposed an
entropy arising from quantum mechanics and we have investigated
its relation to Verlinde's theory. This was then applied to
generalize the Newtonian potential arising in Verlinde's theory.
There are many open problems remaining as this is a theory in
progress. The study of multiple particle situations in the quantum
mechanical approach should be interesting. A relativistic approach
to quantum entropic gravity also needs to be established and
investigated. The construction of various quantum field theories
in curved spacetimes based on this approach should be of
particular interest. Generally the connection to quantized gravity
theories, if there are valid such \cite{Jacobsson}, is left for
further investigation. The nature of space and time as derived
concepts, as spoken of by Verlinde \cite{Verlinde}, is not
addressed in this paper and are features in need of investigation.
The relation to AdS/CFT correspondence is also left open for
further investigation. In conclusion, this paper, guided by a pure
speculation, suggests that the gravitational attraction perhaps
could be the result of a particular type of entropy arising in
quantum mechanics.


\subsection{Final comments}
Verlinde proposed a theory of gravity where gravity no longer was
a fundamental force, but rather an effect of entropy. His theory,
as a reversal of physics research, is a remarkable framework that
has many open questions and interesting consequences still to be
uncovered. This is why the greatest achievement of this paper is
that it provides Verlinde's remarkable theory with a possible
physical explanation for the factor $\hbar$, which had previously
been added by Bekenstein mainly for dimensional reasons.




%



\section{Acknowledgements}
I would like to thank prof. Erik Verlinde for his comments on this
paper.

\end{document}